# Spin wave nonreciprocity for logic device applications

Mahdi Jamali[1], Jae Hyun Kwon[1], Soo-Man Seo[2], Kyung-Jin Lee[2,3], and Hyunsoo Yang[1,*]

[1] Department of Electrical and Computer Engineering, National University of Singapore, 4 Engineering Drive 3, 117576, Singapore

[2] Department of Materials Science and Engineering, Korea University, Seoul 136-701, Korea

[3] KU-KIST Graduate School of Converging Science and Technology, Korea University, Seoul 136-713, Korea

*Correspondence and requests for materials should be addressed to H.Y. (eleyang@nus.edu.sg)

**The utilization of spin waves as eigenmodes of the magnetization dynamics for information processing and communication has been widely explored recently due to its high operational speed with low power consumption and possible applications for quantum computations. Previous proposals of spin wave Mach-Zehnder devices were based on the spin wave phase, a delicate entity which can be easily disrupted. Here, we propose a complete logic system based on the spin wave amplitude utilizing the nonreciprocal spin wave behavior excited by microstrip antennas. The experimental data reveal that the nonreciprocity of magnetostatic surface spin wave can be tuned by the bias magnetic field. Furthermore, engineering of the device structure could result in a high nonreciprocity factor for spin wave logic applications.**

Since the scaling of the conventional semiconductor devices approaches its physical and operational limit[1], various schemes have been proposed to improve the performance of devices in terms of the computational capability and power dissipation including the utilization of photon[2,3], phonon[4], and spin of electrons[5-10]. Spintronics is an emerging technology which exploits the intrinsic spin of electrons and its associated magnetic moments, in addition to their fundamental electronic charges. Different spin based logic circuits have been proposed including magnetic cellular automata[11,12], programmable magnetoresistive elements[13], domain walls[14-16], spin currents[17,18], and spin wave[19,20] based logic devices. Spin waves have a characteristic frequency in the gigahertz range[21-25] with a large group velocity (~ a few tens of μm/ns), and the capability to be guided along a ferromagnetic material allows it to be considered as a promising candidate for information processing[26-29]. The previous studies of the spin wave logic devices based on the phase of spin wave require a controlled phase shifter which complicates the device structure[19,30,31] especially for submicron size devices where the lateral confinement forces the magnetic system to obtain more than one spin wave frequency with different phase velocities[21]. In addition, the spin wave phase is a continuous variable and delicate entity that could be easily disrupted by imperfections and magnetic inhomogeneities[32], and is extremely sensitive to the dimensions of devices, which makes the spin wave logic devices to be less reliable and high bit error rates[33].

In contrast to a widely used dielectric material, yttrium iron garnet (YIG) in the spin wave studies, metallic ferromagnetic materials are more compatible to the standard semiconductor technology and can be easily scaled down. Depending on the relative orientation between the external magnetic field and the spin wave propagation direction, the different modes of magnetostatic spin waves can exist. When the magnetic field is in the plane of the ferromagnetic film and normal to the wavevector of spin waves, the Damon-Eshbach (DE) surface spin wave propagates on the film surfaces[34,35]. It has been known that the surface mode has a nonreciprocal behavior in the spin wave amplitude for the spin waves with opposite signs of wavevectors ($\pm k$).[34-37] In thin ferromagnetic films, the nonreciprocal behavior of magnetostatic surface spin waves is a consequence of the asymmetric distribution of the out-of-plane component of the excitation field (Supplementary Information S1)[35,38].

Here we show that the nonreciprocity value of surface spin waves is not a constant, but it can be tuned by the external bias magnetic field. We further discuss how to engineer a larger nonreciprocity ratio. Utilizing the spin wave amplitude due to the nonreciprocal spin wave behavior, a new type of complete spin wave logic device is proposed.

**Results**

**Nonreciprocal spin wave propagation**. The schematic diagram of the device structure is shown in Fig. 1(a) consisting of a 20 nm thick $Ni_{80}Fe_{20}$ (permalloy) film isolated from the excitation and detection striplines (antennas) by an oxide layer. The bias magnetic field is applied in the $y$-direction and the propagating spin wave in the $x$-direction is detected using an inductive technique. The optical image and scanning electron micrograph (SEM) of the fabricated device are shown in Fig. 1(b, c), respectively, in which $g = 5$ μm is the gap distance between the excitation and detection stripline, $W = 9$ μm is the width of the ground line, and $S = 3$ μm is the width of the signal line. Two different characterization methods have been performed such as frequency domain measurements using a sinusoidal microwave excitation and time domain measurements using a pulse inductive microwave magnetometer.

The frequency spectra of the propagating spin wave for magnetic fields of ±135, ±225, and ±300 Oe are shown in Fig. 2(a). One can see a clear nonreciprocity of the surface spin wave for positive and negative fields which is equivalent to the opposite signs of wavevectors (Supplementary Information S2). Furthermore, spin waves contain two major frequencies where the amplitude of the higher frequency peak is much smaller than that of the lower frequency one.

The surface spin wave has a dispersion relation of $f_{DE} = \frac{\gamma_0}{2\pi}[H(H+4\pi M_s)+(2\pi M_s)^2(1-e^{-2kd})]^{1/2}$ in which $M_s$ is the saturation magnetization, and $\gamma_0$ is the gyromagnetic constant, $d$ is the film thickness, and $k$ is the spin wave wavevector[34]. The spin wave spectrum that is measured for different magnetic fields is shown in Fig. 2(b). A clear quadratic-like behavior of the spin wave frequency versus magnetic field is consistent with the surface spin wave dispersion relationship. By fitting the main frequency of the experimental data in Fig. 2(b) with the spin wave dispersion relation, a wavevector of $k \pm \Delta k = 0.57 \pm 0.33$ μm$^{-1}$ is obtained which is in line with the previous reports with inductive method characterization[25,37]. The observation of two frequencies for the spin wave at a bias magnetic field can be understood based on the presence of two different wavevectors. $\Delta k$ is proportional to the inverse of the signal line width ($S$),[25] therefore for a narrow stripline (3 μm), the different wavevectors are present, while only one frequency is present for a wide stripline (10 μm) (Supplementary Information S3). In Fig. 2(b) the lower branch corresponding to positive bias fields has larger amplitudes compared to the upper branch with negative bias fields, demonstrating the nonreciprocity of spin waves.

In order to better understand the nonreciprocity of spin waves, we have also performed micromagnetic simulations as can be seen in Fig. 2(c) (Supplementary Information S4). The simulation results are similar to the experimental results. In Fig. 2(d), we have calculated the spin wave nonreciprocity factor $NR = \frac{A(H_+)}{A(H_-)} = \frac{A(k_+)}{A(k_-)}$ in the frequency domain which is the ratio of the spin wave amplitude at opposite bias magnetic fields or at opposite wavevectors. The spin wave nonreciprocity is a function of the bias field and it increases upon increasing the bias magnetic field. The nonreciprocity factor is similar for an impulse and microwave sinusoidal excitations[36,37] (Supplementary Information S5). The simulated nonreciprocity matches well with the experimental results as shown in Fig. 2(d).

**Nonreciprocity of spin waves in time domain**. We have also performed time-resolved measurements of spin wave dynamics in Fig. 3(a-c) for a magnetic field of ±60, ±135, and ±200 Oe, respectively. It is clear that the nonreciprocity of spin wave has increased by increasing the field from 60 to 200 Oe which is also consistent with the frequency domain data in Fig. 2(d). The micromagnetic simulation results in Fig. 3(d-f) also show an increase in the nonreciprocity by increasing the bias magnetic field similar to the experimental results in Fig. 3(a-c). Note that the detected spin wave voltage is associated to the change of the magnetization in time. The above experimental and simulation results show that the amplitude of spin-wave packet can be changed by inversing the magnetization direction or by inversing the propagation direction. Therefore, the nonreciprocity of the surface spin wave amplitude is a promising candidate for the implementation of spin wave logic circuits.

**Logic gates using nonreciprocal spin wave propagation**. Here, we propose spin wave logic gates to implement any Boolean function. A cross-section image of a NOT gate or a PASS gate can be seen in Fig. 4(a) and the top view image of the device structure is shown in Fig. 4(b). We assume that the ferromagnetic structure has an anisotropy with an easy-axis in the *y*-direction, which can be easily implemented by material engineering during the film growth[39,40]. Once the input *A* is a logic 1, the Oersted field generated by the input *A*-line is strong enough to switch the magnetization of the ferromagnetic film to the –y-direction. When the input *A* is a logic 0, the Oersted field generated by the input *A*-line favors the magnetization to be aligned in the +*y*-direction. The device has two outputs ($Y$ and $\overline{Y}$) complementary to each other. If the effective field is in the +y-direction, the spin wave excited by the narrow current pulse of triggering signal *I* generates a larger amplitude in $Y$ than $\overline{Y}$. For an effective field in the –y-direction, the output $\overline{Y}$ has larger amplitudes than $Y$. The truth table of the logic circuit is shown in Fig. 4(c). The output $Y$ implements a NOT gate, while the $\overline{Y}$ implements a PASS gate.

We can also implement a two-input logic circuit as shown in Fig. 4(d-e). The magnetic film has an easy-axis in the y-direction with an anisotropy field of $H_b$. The input $A$ generates a field large enough to switch the ferromagnetic film in the $-y$-direction, while the input B generates a field in the $+y$-direction. The truth-table of the logic circuit is shown in Fig. 4(f) with corresponding schematics of logic gates. The Boolean expression of the output $Y$ can be written as $Y = \overline{A} + B$, which is a combination of NOT and OR gates. We can build all the logic gates with the help of the proposed one-input and two-input gates (Supplementary Information S6). Therefore, the proposed logic gates are a complete logic system to build any Boolean function. For a short pulse excitation in the triggering signal $I$, the output signal is a spin wave packet that can be reshaped using a standard push-detection circuit commonly used in the *rf* systems, therefore, the output signal can be fed as an input for the other gates (Supplementary Information S7).

**Discussion**

The triggering signal $I$ can be used as a clock for the synchronization of different gates, therefore depending on the propagation delay of the spin wave and the delay of the wave-shaping circuit, the maximum operational frequency of the logic gate can be determined. The surface spin wave has a very large group velocity ($v_g \sim$ a few tens of μm/ns)[23]. Furthermore, the push-detector circuit can operate in the GHz range (See for example LMH2110 power detector chip). Therefore, the logic gate can easily operate in the GHz frequency ranges. In addition, the device is scalable and the performance of the device can be even enhanced in low dimensions (Supplementary Information S8). We have designed logic gates with a feature size of 10 and 100 nm for a width of the ferromagnetic structure as an example of the scalability of the device (Supplementary Information S9). In terms of power consumption, the spin wave propagation does not involve any charge transfer, therefore it is an energy efficient logic scheme. The triggering signal *I* can be a very short pulse (< 80 ps) as it is demonstrated in our experiments, and the driving current of *I* can be very small (< 1 mA) since the field induced by *I* is inversely proportional to the width of the strip line. In addition, our spin wave logic has some degrees of non-volatility such that the magnetization direction of the ferromagnetic structure is preserved in the ±*y* directions, when the device power is turned off.

In summary, the nonreciprocal behavior of spin waves has been experimentally studied in both frequency and time domains. It is found that the nonreciprocity factor is not constant and can be tuned by the bias magnetic field. Based on the findings, a new type of spin wave based logic devices has been proposed that utilizes the spin wave amplitude for its operations. We show that our logic system is complete and can implement all the Boolean functions. Our proposed logic circuit operates in GHz ranges with very low power dissipation, and paves a way to the future spin-based logic devices.

**Methods**

**Device fabrication**. A 20 nm thick permalloy film is dc-sputtered at room temperature on Si/SiO$_2$ (300 nm) substrates. After coating the substrate with a negative-tone e-beam resist, the resist are patterned into different structures with a length of 200 μm and the width ranging from 1 to 400 μm. In the case of small wires (width < 11 μm), multiple wires with a spacing of 1 μm and a total width of 400 μm are patterned. The patterns are transferred to the ferromagnetic film using Ar ion milling and then the resist is removed. Using ac-magnetron sputtering, a 50 nm thick SiO$_2$ layer is deposited for the isolation of the ferromagnetic structure from the waveguides. Using the second e-beam lithography step with a positive-tone resist, the asymmetric coplanar waveguides are patterned followed by the deposition of Cr (5nm)/Ag (150 nm)/Pt (5 nm) and lift-off.

**Measurements**. The samples are characterized in a homemade high frequency probe station with *in-situ* magnetic fields (up to 6 kOe) and using GSG probes from GGB Industries, Inc. For the frequency-domain characterization of spin waves, we utilize an Agilent N5245A microwave network analyzer with an excitation power of 5 dBm. Before frequency-domain characterization, the effects of cables and connectors are compensated using a CS-5 calibration substrate from GGB Industries, Inc. For time-domain characterization, we utilize an 86100A Infiniium wide-bandwidth sampling oscilloscope. The pulses are generated using a Centellax PPG12500 pulse/pattern generator with an amplitude of 1.6 V. Before capturing of spin waves, the signal is amplified by 29 dB using a SHF810 low noise preamplifier.

**Acknowledgements**

This work is partially supported by the Singapore Ministry of Education Academic Research Fund Tier 1 (R-263-000-A46-112) and Singapore National Research Foundation under CRP Award No. NRF-CRP 4-2008-06.

**Author contributions**

M.J. and H.Y. conceived the experiments. M.J. and J.K. carried out experiments. M.J., S.S., and K.L. did simulations. M.J. and H.Y. wrote the manuscript. All authors discussed the data and the results, and commented the manuscript.

**Additional information**

**Supplementary Information** accompanies this paper on www.nature.com/scientificreports

**Competing financial interests:** The authors declare no competing financial interests.

**Figure 1: Device structure**. (a) Schematic illustration of the device structure. The device includes a 20 nm thick permalloy film separated from the striplines by 50 nm of $SiO_2$. The striplines are in an asymmetric coplanar waveguide configuration made of Cr (5 nm)/Ag (150 nm)/Pt (5 nm) with a signal width (*S*) of 3 μm and ground width (*W*) of 9 μm. The bias magnetic field is applied in the *y*-direction and propagating spin wave in the *x*-direction is detected by the antenna using an inductive technique. (b) The optical microscope picture of a device. (c) Scanning electron micrograph (SEM) of a device.

**Figure 2: Nonreciprocal spin wave propagation in frequency domain**. (a) Spin wave excitation spectra measured using a vector network analyzer for different magnetic fields of ±135, ±225, and ±300 Oe. The data of the positive fields are shifted by 0.01 dBm for clarity. (b) Spin wave frequency measured at different magnetic fields. (c) Simulation results of the surface spin wave at different magnetic fields. (d) Nonreciprocity factors of surface spin wave measured at different magnetic fields in both frequency and time domains, and the corresponding micromagnetic simulation results.

**Figure 3: Time domain characterization of nonreciprocity of spin waves**. The time resolved measurement of surface spin waves for three bias magnetic fields of ±60 Oe (a), ±135 Oe (b), and ±200 Oe (c). An excitation impulse voltage of 1.6 V with a pulse width of ~ 80 ps has been used. Micromagnetic simulation results of surface spin waves at ±60 Oe (d), ±135 Oe (e), and ±200 Oe (f).

**Figure 4: Logic gates using nonreciprocal spin wave propagation**. The spin wave logic based on the spin wave nonreciprocal behavior. The schematic diagram of the cross section view (a) and the top view (b) of logic device structure for one input (*A*) and two complementary outputs

($Y$ and $\overline{Y}$). (c) The truth table of the logic gate with the corresponding Boolean expression for each output that resembles a NOT gate for the $Y$ output and a PASS gate for the $\overline{Y}$ port. The schematic diagram of the cross section view (d) and the top view (e) of the device structure for two-input ($A$ and $B$) logic gate. (f) The truth table and Boolean expressions of the two-input logic gate.

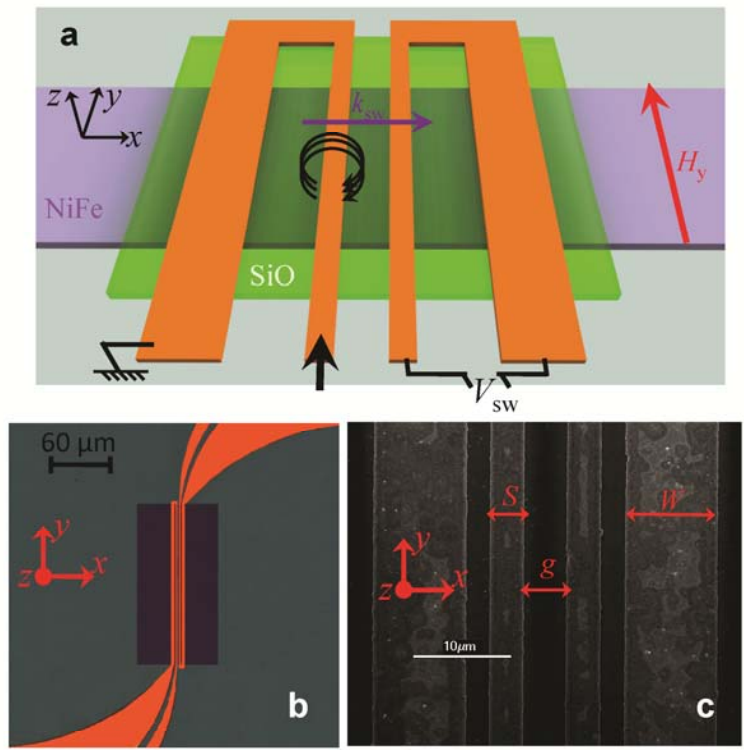

Fig. 1

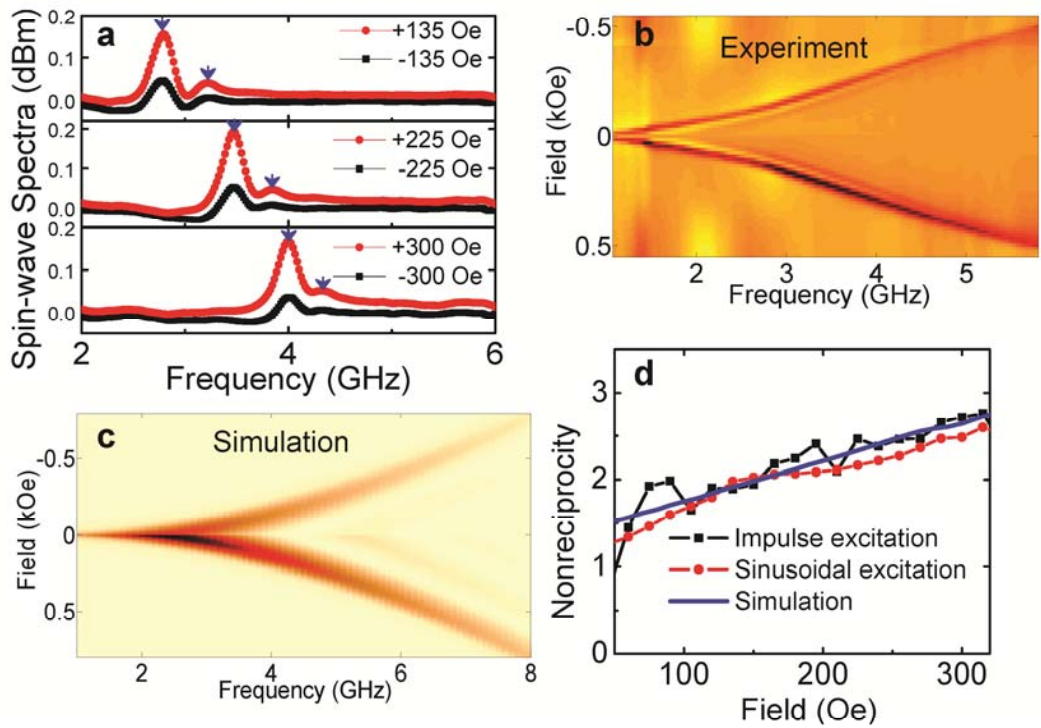

Fig. 2

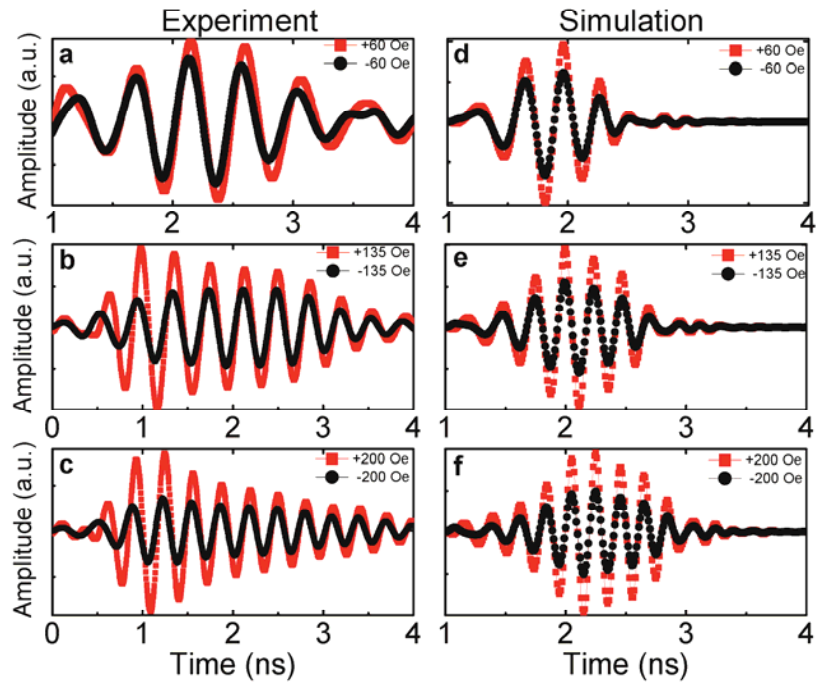

Fig. 3

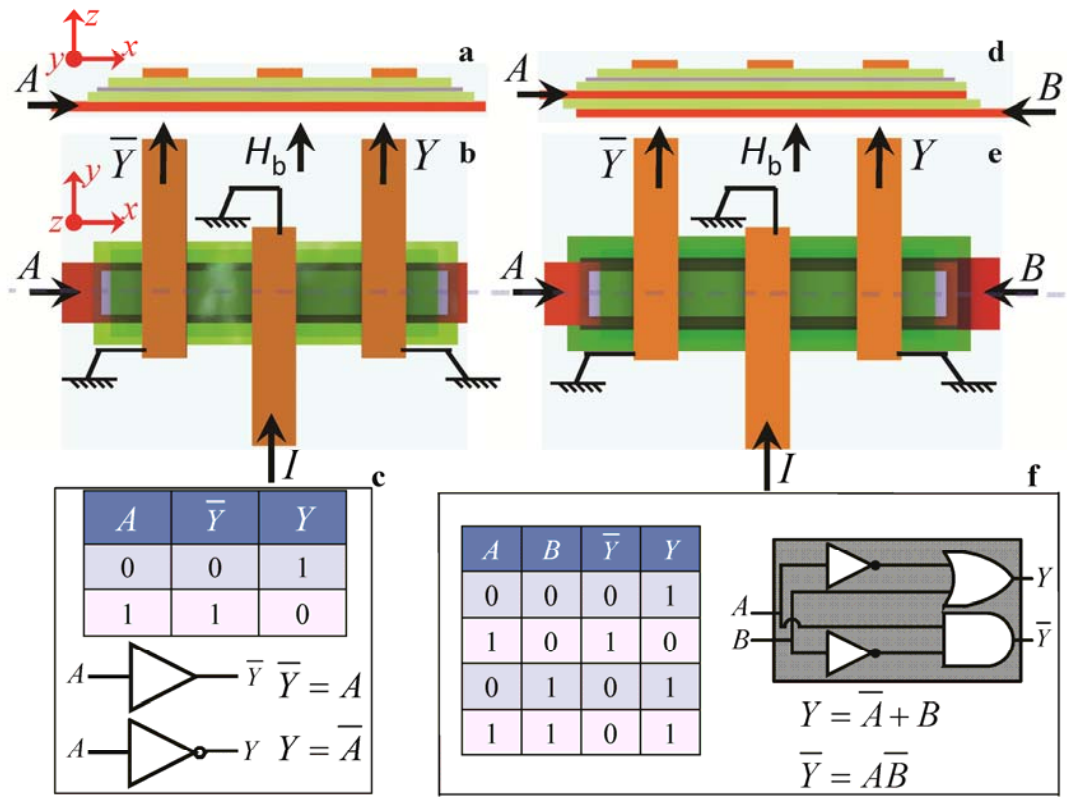

Fig. 4